\documentclass[DIV12]{scrartcl}

\usepackage[dvipsnames,table]{xcolor}
\usepackage{hyperref}
\usepackage[utf8]{inputenc}
\usepackage{tabularx}
\usepackage{setspace}
\usepackage{graphicx}
\usepackage{booktabs}
\usepackage{amsmath}
\usepackage{numprint} 
\usepackage{url}
\usepackage{ragged2e}
\usepackage{todonotes}
\usepackage[inline]{enumitem}
\selectcolormodel{gray}

\hypersetup{
	colorlinks=true,
	linkcolor=black,
	citecolor=black,
	urlcolor=black
}

\date{}
\usepackage{authblk}
\usepackage{blindtext}
\usepackage{natbib}
\title{Chord Label Personalization through Deep \\
Learning of Integrated Harmonic Interval-based 
Representations}

\usepackage{fancyhdr}
\begin{document}

\newcommand{\ace}{\textsc{ace}}
\newcommand{\cqt}{\textsc{cqt}}
\newcommand{\dnn}{\textsc{dnn}}
\newcommand{\hic}{\textsc{hip}}
\newcommand{\ahic}{\textsc{ship}}
\newcommand{\gt}{\textsc{iso}}
\newcommand{\mirex}{\textsc{mirex}}
\newcommand{\comp}{$\left.\textsc{ann}\right|_{\gt{}}$}
\newcommand{\note}[1]{\colorbox{BurntOrange}{Note: #1}}
\newcommand{\ccell}[1]{\cellcolor[gray]{#1} \large #1}
\newcommand{\hipl}{Harmonic Interval Profiles}
\newcommand{\hip}{\textsc{hip}}
\newcommand{\dnns}[1]{$\textsc{dnn}_{#1}$}

\author[1]{Hendrik Vincent Koops\thanks{h.v.koops@uu.nl}}
\author[2]{W. Bas de Haas\thanks{bas@chordify.net}}
\author[2]{Jeroen Bransen\thanks{jeroen@chordify.net}}
\author[1]{Anja Volk\thanks{a.volk@uu.nl} }

\affil[1]{\small Utrecht University, Utrecht, the Netherlands}
\affil[2]{\small Chordify, Utrecht, the Netherlands}

\maketitle
\thispagestyle{fancy} œ

\begin{abstract}
\noindent 
The increasing accuracy of automatic chord estimation systems, the
availability of vast amounts of heterogeneous reference annotations, and
insights from annotator subjectivity research make chord label personalization
increasingly important. Nevertheless, automatic chord estimation systems are
historically exclusively trained and evaluated on a single reference
annotation. We introduce a first approach to automatic chord label
personalization by modeling subjectivity through deep learning of a harmonic
interval-based chord label representation. After integrating these
representations from multiple annotators, we can accurately personalize chord
labels for individual annotators from a single model and the annotators' chord
label vocabulary. Furthermore, we show that chord personalization using
multiple reference annotations outperforms using a single reference
annotation.

\bigskip


\end{abstract}

\section{Introduction}
Annotator subjectivity makes it hard to derive one-size-fits-all chord labels.
Annotators transcribing chords from a recording by ear can disagree because
of personal preference, bias towards a particular instrument, and because
harmony can be ambiguous perceptually as well as theoretically by definition
\citep{schoenberg1978theory,meyer1957meaning}. These reasons contributed to
annotators creating large amounts of heterogeneous chord label reference
annotations. For example, on-line repositories
for popular songs often contain multiple, heterogeneous versions.

One approach to the problem of finding the appropriate chord labels in a large
number of heterogeneous chord label sequences for the same song is data fusion.
Data fusion research shows that knowledge shared between sources can be
integrated to produce a unified view that can outperform individual sources
\citep{dong2009integrating}. In a musical application, it was found that
integrating the output of multiple Automatic Chord Estimation (\ace{})
algorithms results in chord label sequences that outperform the individual
sequences when compared to a single ground truth \citep{koopsintegration}.
Nevertheless, this approach is built on the intuition that one single correct
annotation exists that is best for everybody, on which
\ace{} systems are almost exclusively trained.
Such reference annotation is either compiled by a single person
\citep{mauch2009omras2}, or unified from multiple opinions
\citep{burgoyne2011expert}. Although most of the creators of these datasets
warn for subjectivity and ambiguity, they are in
practice used as the \emph{de facto} ground truth in MIR chord research and
tasks (e.g. \mirex{} \ace{}).

On the other hand, it can also be argued that there is no single best reference
annotation, and that chord labels are correct with varying degrees of
{``goodness-of-fit''} depending on the target audience
\citep{ni2013understanding}. In particular for richly orchestrated,
harmonically complex music, different chord labels can be chosen for a
part, depending on the instrument, voicing or the annotators' chord label
vocabulary.

In this paper, we propose a solution to the problem of finding appropriate chord
labels in multiple, subjective heterogeneous reference annotations for the same
song. We propose an automatic audio chord label estimation and personalization
technique using the harmonic content shared between annotators. From
deep learned shared harmonic interval profiles, we can create chord labels that
match a particular
\emph{annotator vocabulary}, thereby providing an annotator with familiar, and
personal chord labels. We test our approach on a 20-song dataset with
multiple reference annotations, created by annotators who use different chord
label vocabularies. We show that by taking into account annotator subjectivity
while training our \ace{} model, we can provide personalized chord labels for
each annotator.

\textbf{Contribution.} The contribution of this paper is twofold.
First, we introduce an approach to automatic chord label personalization by 
taking into account annotator subjectivity. Through this end, we introduce a
harmonic interval-based mid-level representation that captures harmonic 
intervals found in chord labels. 
Secondly, we show that after integrating these features from multiple annotators
and deep learning, we can accurately personalize chord labels for individual 
annotators. Finally, we show that chord label personalization using integrated 
features outperforms personalization from a commonly used reference annotation.


\section{Deep Learning Harmonic Interval Subjectivity}
\label{sec:method}
For the goal of chord label personalization, we create an harmonic
bird's-eye view from different reference annotations, by integrating their chord
labels. More specifically, we introduce a new feature that captures the
shared harmonic interval profile of multiple chord labels, which we 
deep learn from audio. 
First, we extract Constant Q (\cqt{}) features from audio, then we calculate
Shared Harmonic Interval Profile (\ahic{}) representations from multiple chord
label reference annotations corresponding to the \cqt{} frames. Finally, we
train a deep neural network to associate a context window of \cqt{} to \ahic{}
features.

From audio, we calculate a time-frequency representation where the frequency
bins are geometrically spaced and ratios of the center frequencies to bandwidths
of all bins are equal, called a Constant Q (\cqt{}) spectral transform
\citep{schorkhuber2010constant}. 
We calculate these \cqt{} features with a hop length of $4096$ samples, a
minimum frequency of $\approx 32.7$ Hz (C1 note), $24 \times 8 = 192$ bins, $24$
bins per octave. This way we can capture pitches spanning from low notes to 8
octaves above C1. Two bins per semitone allows for slight tuning variations.

\begin{table}
\centering
\resizebox{\textwidth}{!}{%
\begin{tabular}{lccccccccccccccccccc}
& C  &C\# &D  &D\# &E  &F  &F\# &G  &G\# & A &A\# &B & N & $\sharp$3
& $\flat$3 & $\star3$ & $\sharp$7 & $\flat$7 & $\star7$ \\ \hline 
G:maj7 & 0 & 0 & 0 & 0 & 0 & 0 & 0 & 1 & 0 & 0 & 0 & 0 & 0 & 1 & 0 & 0 & 1 & 0 
& 0 \\ 
G:maj & 0 & 0 & 0 & 0 & 0 & 0 & 0 & 1 & 0 & 0 & 0 & 0 & 0 & 1 & 0 & 0 & 0 & 0 
& 1 \\ 
G:maj7 & 0 & 0 & 0 & 0 & 0 & 0 & 0 & 1 & 0 & 0 & 0 & 0 & 0 & 1 & 0 & 0 & 1 & 0 
& 0 \\ 
G:minmaj7 & 0 & 0  & 0 & 0 & 0  & 0 & 0 & 1 & 0 & 0  & 0 & 0 & 0 & 0 & 1 & 0 
& 1 & 0 & 0 \\ \hline
\ahic{} & 0 & 0 & 0 & 0 & 0 & 0 & 0 & 1 & 0 & 0 & 0 & 0 & 0 & 0.75 & 0.25 & 0 
& 0.75 & 0 & 0.25 \\
\end{tabular}
}
\caption{Interval profiles from root notes of \hic{}s of different chord labels 
and their \ahic{}}
\label{tab:hierarchy}
\end{table}
To personalize chord labels from an arbitrarily sized vocabulary for an
arbitrary number of annotators, we need a chord representation that
\begin{enumerate*}[label=(\roman*)]
  \item is robust against label sparsity, and
  \item captures an integrated view of all annotators.
\end{enumerate*}
We propose a new representation that captures a harmonic interval profile
(\hic{}) of chord labels, instead of directly learning a chord label classifier.
The rationale behind the \hic{} is that most chords can be reduced to the root
note and the stacked triadic intervals, where the amount and combination of
triadic interval determines the chord quality and possible extensions. The
\hic{} captures this intuition by reducing a chord label to its root and
harmonic interval profile. \hic{} is a concatenation of multiple one-hot vectors
that denote a root note and additional harmonic intervals relative to the root
that are expressed in the chord label.

In this paper, we use a concatenation of three one-hot vectors: roots, thirds
and sevenths. The first vector is of size 13 and denotes the 12 chromatic root
notes (C\dots B) + a ``no chord'' (N) bin. The second vector is of size 3 and
denotes if the chord denoted by the chord label contains a major third
($\sharp3$), minor third ($\flat3$), or no third ($\star3$) relative to the root
note. The third vector, also of size 3, denotes the same, but for the seventh
interval ($\sharp7,\flat7,\star7$). The \hic{} can be extended to include other
intervals as well. In Table \ref{tab:hierarchy} we show example chord labels and
their \hic{} equivalent. The last row shows the \ahic{} created from the \hic{}s
above it.

\subsection{Deep Learning Shared Harmonic Interval Profiles}
\label{sec:avgchorma}
We use a deep neural network to learn \ahic{} from \cqt{}. Based on preliminary
experiments, a funnel-shaped architecture with three hidden rectifier unit
layers of sizes 1024, 512, and 256 is chosen. Research in audio content analysis
has shown that better prediction accuracies can be achieved by aggregating
information over several frames instead of using a single frame
\citep{sigtia2015audio,bergstra2006aggregate}. Therefore, the input for our
\dnn{} is a window of \cqt{} features from which we learn the \ahic{}.
Preliminary experiments found an optimal window size of 15 frames, that is: 7
frames left and right directly adjacent to a frame. Consequently, our neural
network has input layer size of $192\times15=2880$. The output layer consists of
19 units corresponding with the \ahic{} features as explained above.

We train the \dnn{} using stochastic gradient descent by minimizing the cross-
entropy between the output of the \dnn{} with the desired \ahic{} (computed by
considering the chord labels from all annotators for that audio frame). We
train the hyper-parameters of the network using mini-batch (size 512) training
using the \textsc{adam} update rule \citep{kingma2014adam}. Early stopping is
applied when validation accuracy does not increase after 20 epochs. After
training the \dnn{}, we can create chord labels from the learned \ahic{}
features.


\section{Annotator Vocabulary-based Chord Label Estimation}
\label{sec:testcase1}
The \ahic{} features are used to associate probabilities to chord labels from a
given vocabulary. For a chord label $\mathbf{L}$ the \hic{} $\mathbf{h}$
contains exactly three ones, corresponding to the root, thirds and sevenths of
the label $\mathbf{L}$. From the \ahic{} $\mathbf{A}$ of a particular audio
frame, we project out three values for which $\mathbf{h}$ contains ones
($\mathbf{h}(\mathbf{A})$). The product of these values is then interpreted as
the combined probability $\mathbf{CP}$ ($= \Pi\ \mathbf{h}(\mathbf{A})$) of the
intervals in $\mathbf{L}$ given $\mathbf{A}$. Given a vocabulary of chord
labels, we normalize the $\mathbf{CP}$s to obtain a probability density function
over all chord labels in the vocabulary given $\mathbf{A}$. The chord label with
the highest probability is chosen as the chord label for the audio frame
associated to $\mathbf{A}$.

For the chord label examples in Table \ref{tab:hierarchy}, the products of the
non-zero values of the point-wise multiplications $\approx$ 0.56, 0.19, and 0.19
for G:maj7, G:maj, and G:minmaj7 respectively. If we consider these chord labels
to be a vocabulary, and normalize the values, we obtain probabilities $\approx$
0.6,  0.2,  0.2, respectively. Given extracted \ahic{} from multiple annotators
providing reference annotations and chord label vocabularies, we can now
generate annotator specific chords labels.

\section{Evaluation}
\label{sec:evaluation}
\ahic{} models multiple (related) chords for a single frame, e.g.,
the \ahic{} in Table 1 models different flavors of a G and a C chord. For the
purpose of personalization, we want to present the annotator with only the
chords they understand and prefer, thereby producing a high chord label accuracy
for each annotator. For example, if an annotator does not know a G:maj7 but does
know an G, and both are probable from an \ahic{}, we like to present the latter.
In this paper, we evaluate our \dnn{} \ace{} personalization approach, and the
\ahic{} representation, for each individual annotator and their vocabulary.

In an experiment we compare training of our chord label personalization system
on multiple reference annotations with training on a commonly used single
reference annotation. In the first case we train a \dnn{} (\dnns{\ahic{}}) on
\ahic{}s derived from a dataset introduced by \cite{ni2013understanding}
containing 20 popular songs annotated by five annotators with varying degrees of
musical proficiency. In the second case, we train a \dnn{} (\dnns{\gt{}}) on the
\hic{} of the \emph{Isophonics} (\gt{}) single reference annotation
\citep{mauch2009omras2}. \gt{} is a peer-reviewed, and \emph{de facto} standard 
training reference annotation used in numerous \ace{} systems. From the
(s)\hic{} the annotator chord labels are derived and we evaluate the systems on
every individual annotator. We hypothesize that training a system on \ahic{}
based on multiple reference annotations captures the annotator subjectivity of
these annotations and leads to better personalization than training the same
system on a single (\gt{}) reference annotation.

It could be argued that the system trained on five reference annotations has
more data to learn from than a system trained on the single \gt{} reference
annotation. To eliminate this possible training bias, we evaluate the
annotators' chord labels directly on the chord labels from \gt{} (\comp). This
evaluation reveals the similarity between the \ahic{} and the \gt{} and puts the
results from \dnns{\gt{}} in perspective. If \dnns{\ahic{}} is better at
personalizing chords (i.e. provides chord labels with a higher accuracy per
annotator) than \dnns{\gt{}} while the annotator's annotations and the \gt{} are
similar, then we can argue that using multiple reference annotations and \ahic{}
is better for chord label personalization than using just the \gt{}. In a final
baseline evaluation, we also test \gt{} on \dnns{\gt{}} to measure how well it
models the \gt{}.

Ignoring inversions, the complete dataset from \cite{ni2013understanding}
contains 161 unique chord labels, comprised of five annotators using 87, 74, 62,
81 and 26 unique chord labels respectively. The intersection of the chord labels
of all annotators contains just 21 chord labels meaning that each annotator uses
a quite distinct vocabulary of chord labels. For each song in the dataset, we
calculate \cqt{} and \ahic{} features. We divide our \cqt{} and \ahic{} dataset
frame-wise into 65\% training (28158 frames), 10\% evaluation (4332 frames) and 
25\% testing (10830 frames) sets.
For the testing set, for each annotator, we create chord labels from the
deep learned \ahic{} based on the annotators' vocabulary.
 
We use the standard \textsc{mirex} chord label evaluation methods to compare the
output of our system with the reference annotation from an annotator
\citep{raffel2014mir_eval}. We use evaluations at different chord granularity
levels. \textsc{root} only compares the root of the chords.
\textsc{majmin} only compares major, minor, and “no chord” labels.
\textsc{mirex} considers a chord label correct if it shares at least
three pitch classes with the reference label.
\textsc{thirds} compares chords at the level of root and major or minor third. 
\textsc{7ths} compares all above plus the seventh notes.

\begin{table}
\centering
 \resizebox{\textwidth}{!}{%
\begin{tabular}{lccc|ccc|ccc|ccc|ccc|c}
~ & \multicolumn{3}{c}{Annotator 1} & \multicolumn{3}{c}{Annotator 2} &
  \multicolumn{3}{c}{Annotator 3} & \multicolumn{3}{c}{Annotator 4} &
  \multicolumn{3}{c|}{Annotator 5} & \gt{} \\ 
~ & \dnns{\ahic{}}{} & \comp{} & \dnns{\gt{}} &
  \dnns{\ahic{}}{} & \comp{} & \dnns{\gt{}} & 
  \dnns{\ahic{}}{} & \comp{} & \dnns{\gt{}} & 
  \dnns{\ahic{}}{} & \comp{} & \dnns{\gt{}} & 
  \dnns{\ahic{}}{} & \comp{} & \dnns{\gt{}} & 
  \dnns{\gt{}} \\ \hline
\textsc{root}&\ccell{0.85}&\ccell{0.73}&\ccell{0.66}&\ccell{0.82}&\ccell{0.74}
&\ccell{0.67}&\ccell{0.80}&\ccell{0.72}&\ccell{0.65}&\ccell{0.80}&\ccell{0.73}
&\ccell{0.65}&\ccell{0.77}&\ccell{0.67}&\ccell{0.60}&{0.86}\\
\textsc{majmin}&\ccell{0.82}&\ccell{0.69}&\ccell{0.61}&\ccell{0.69}&\ccell{0.67}
&\ccell{0.53}&\ccell{0.67}&\ccell{0.69}&\ccell{0.53}&\ccell{0.73}&\ccell{0.67}
&\ccell{0.55}&\ccell{0.72}&\ccell{0.61}&\ccell{0.55}&{0.69}\\
\textsc{mirex}&\ccell{0.82}&\ccell{0.70}&\ccell{0.61}&\ccell{0.69}&\ccell{0.68}
&\ccell{0.54}&\ccell{0.66}&\ccell{0.69}&\ccell{0.54}&\ccell{0.73}&\ccell{0.68}
&\ccell{0.56}&\ccell{0.72}&\ccell{0.62}&\ccell{0.55}&{0.69}\\
\textsc{thirds}&\ccell{0.82}&\ccell{0.70}&\ccell{0.62}&\ccell{0.75}&\ccell{0.67}
&\ccell{0.59}&\ccell{0.79}&\ccell{0.69}&\ccell{0.62}&\ccell{0.76}&\ccell{0.68}
&\ccell{0.61}&\ccell{0.72}&\ccell{0.62}&\ccell{0.55}&{0.83}\\
\textsc{7ths}&\ccell{0.77}&\ccell{0.56}&\ccell{0.50}&\ccell{0.64}
&\ccell{0.53}&\ccell{0.42}&\ccell{0.64}&\ccell{0.56}&\ccell{0.43}&\ccell{0.53}
&\ccell{0.48}&\ccell{0.40}&\ccell{0.72}&\ccell{0.53}&\ccell{0.55}&{0.65}
\end{tabular}
}
\caption{Chord label personalization accuracies for the five annotators}
\label{tab:results}
\end{table}


\section{Results}
\label{sec:results}

The \dnns{\ahic{}} columns of Table~\ref{tab:results} for each annotator show
average accuracies of 0.72 ($\sigma = 0.08$). For each chord granularity level,
our \dnns{\ahic{}}{} system provides personalized chord labels that are trained
on multiple annotations, but are comparable with a system that was trained an
evaluated on a single reference annotation (\gt{} column of
Tab.~\ref{tab:results}). Comparable high accuracy scores for each annotator show
that the system is able to learn a \ahic{} representation that
\begin{enumerate*}[label=(\roman*)]
  \item is meaningful for all annotators
  \item from which chord labels can be accurately personalized for each
  annotator.
\end{enumerate*}
The low scores for annotator 4 for \textsc{sevenths} form an exception. An
analysis by \cite{ni2013understanding} revealed that between annotators, 
annotator 4 was on average the most different from the consensus. Equal scores
for annotator 5 for all evaluations except \textsc{root} are explained by
annotator 5 being an amateur musician using only major and minor chords.

Comparing the \dnns{\ahic{}} and \dnns{\gt{}} columns, we see that for each
annotator \dnns{\ahic{}} models the annotator better than \dnns{\gt{}}. With an
average accuracy of 0.55 ($\sigma = 0.07$), \dnns{\gt{}}'s accuracy is on
average 0.17 lower than \dnns{\ahic{}}, showing that for these annotators, \gt{}
is not able to accurately model chord label personalization. Nevertheless, the
last column shows that the system trained on \gt{} modeled the \gt{} quite
well. The results of \comp{} show that the annotators in general agree with
\gt{}, but the lower score in \dnns{\gt{}} shows that the agreement is not good
enough for personalization. Overall, these results show that our system is able
to personalize chord labels from multiple reference annotations, while
personalization using a commonly used single reference annotation yields
significantly worse results.

\section{Conclusions and Discussion}
\label{sec:conclusions} 

We presented a system that provides personalized chord labels
from multiple reference annotations from audio, based on the annotators'
specific chord label vocabulary and an interval-based chord label representation
that captures the shared subjectivity between annotators.
To test the scalability of our system, our experiment needs to be repeated on a
larger dataset, with more songs and more annotators. Furthermore, a similar
experiment on a dataset with instrument/proficiency/cultural-specific
annotations from different annotators would shed light on whether our system
generalizes to providing chord label annotations in different contexts. From the
results presented in this paper, we believe chord label personalization is the
next step in the evolution of \ace{} systems.

\section*{Acknowledgments} 
We thank Y. Ni, M. McVicar, R. Santos-Rodriguez and T. De Bie for
providing their dataset. 

\bibliography{paper}

\begin{thebibliography}{12}
\providecommand{\natexlab}[1]{#1}
\providecommand{\url}[1]{\texttt{#1}}
\expandafter\ifx\csname urlstyle\endcsname\relax
  \providecommand{\doi}[1]{doi: #1}\else
  \providecommand{\doi}{doi: \begingroup \urlstyle{rm}\Url}\fi

\bibitem[Bergstra et~al.(2006)Bergstra, Casagrande, Erhan, Eck, and
  K{\'e}gl]{bergstra2006aggregate}
J.~Bergstra, N.~Casagrande, D.~Erhan, D.~Eck, and B.~K{\'e}gl.
\newblock Aggregate features and adaboost for music classification.
\newblock \emph{Machine learning}, 65\penalty0 (2-3):\penalty0 473--484, 2006.

\bibitem[Burgoyne et~al.(2011)Burgoyne, Wild, and Fujinaga]{burgoyne2011expert}
J.A. Burgoyne, J.~Wild, and I.~Fujinaga.
\newblock An expert ground truth set for audio chord recognition and music
  analysis.
\newblock In \emph{Proc. of the 12th International Society for Music
  Information Retrieval Conference, ISMIR}, volume~11, pages 633--638, 2011.

\bibitem[Dong et~al.(2009)Dong, Berti-Equille, and
  Srivastava]{dong2009integrating}
X.L. Dong, L.~Berti-Equille, and D.~Srivastava.
\newblock Integrating conflicting data: the role of source dependence.
\newblock \emph{Proc. of the VLDB Endowment}, 2\penalty0 (1):\penalty0
  550--561, 2009.

\bibitem[Kingma and Ba(2014)]{kingma2014adam}
D.P. Kingma and J.~Ba.
\newblock Adam: A method for stochastic optimization.
\newblock In \emph{Proc. of the 3rd International Conference on Learning
  Representations, ICLR}, 2014.

\bibitem[Koops et~al.(2016)Koops, de~Haas, Bountouridis, and
  Volk]{koopsintegration}
H.V. Koops, W.B. de~Haas, D.~Bountouridis, and A.~Volk.
\newblock Integration and quality assessment of heterogeneous chord sequences
  using data fusion.
\newblock In \emph{Proc. of the 17th International Society for Music
  Information Retrieval Conference, ISMIR, New York, USA}, pages 178--184,
  2016.

\bibitem[Mauch et~al.(2009)Mauch, Cannam, Davies, Dixon, Harte, Kolozali,
  Tidhar, and Sandler]{mauch2009omras2}
M.~Mauch, C.~Cannam, M.~Davies, S.~Dixon, C.~Harte, S.~Kolozali, D.~Tidhar, and
  M.~Sandler.
\newblock Omras2 metadata project 2009.
\newblock In \emph{Late-breaking demo session at 10th International Society for
  Music Information Retrieval Conference, ISMIR}, 2009.

\bibitem[Meyer(1957)]{meyer1957meaning}
L.B. Meyer.
\newblock Meaning in music and information theory.
\newblock \emph{The Journal of Aesthetics and Art Criticism}, 15\penalty0
  (4):\penalty0 412--424, 1957.

\bibitem[Ni et~al.(2013)Ni, McVicar, Santos-Rodriguez, and
  De~Bie]{ni2013understanding}
Y.~Ni, M.~McVicar, R.~Santos-Rodriguez, and T.~De~Bie.
\newblock Understanding effects of subjectivity in measuring chord estimation
  accuracy.
\newblock \emph{IEEE Transactions on Audio, Speech, and Language Processing},
  21\penalty0 (12):\penalty0 2607--2615, 2013.

\bibitem[Raffel et~al.(2014)Raffel, McFee, Humphrey, Salamon, Nieto, Liang,
  Ellis, and Raffel]{raffel2014mir_eval}
C.~Raffel, B.~McFee, E.J. Humphrey, J.~Salamon, O.~Nieto, D.~Liang, D.P.W.
  Ellis, and C.~Raffel.
\newblock mir\_eval: A transparent implementation of common mir metrics.
\newblock In \emph{Proc. of the 15th International Society for Music
  Information Retrieval Conference, ISMIR}, pages 367--372, 2014.

\bibitem[Schoenberg(1978)]{schoenberg1978theory}
A.~Schoenberg.
\newblock \emph{Theory of harmony}.
\newblock University of California Press, 1978.

\bibitem[Sch{\"o}rkhuber and Klapuri(2010)]{schorkhuber2010constant}
C.~Sch{\"o}rkhuber and A.~Klapuri.
\newblock Constant-q transform toolbox for music processing.
\newblock In \emph{Proc. of the 7th Sound and Music Computing Conference,
  Barcelona, Spain}, 2010.

\bibitem[Sigtia et~al.(2015)Sigtia, Boulanger-Lewandowski, and
  Dixon]{sigtia2015audio}
S.~Sigtia, N.~Boulanger-Lewandowski, and S.~Dixon.
\newblock Audio chord recognition with a hybrid recurrent neural network.
\newblock In \emph{Proc. of the 16th International Society for Music
  Information Retrieval Conference, ISMIR}, pages 127--133, 2015.

\end{thebibliography}
\end{document}